\documentclass[12pt]{article}
\usepackage{graphicx}
\usepackage{amsmath,amssymb,latexsym,epsfig,epsf,rotate}
\def\di{{\mbox{$\bullet$}}}

\begin{document}
\newcommand{\ve }[1]{{\mbox{\boldmath $#1$}}}
\newcommand{\vsm }[1]{{\mbox{\scriptsize {\boldmath $#1$} }}}

\begin{center}
\vspace*{2cm} {\Large Elastic and dynamic form factors of an
atomic nucleus in the shell model with correction for the
center-of-mass motion\footnote{Paper published in Ukr.J.Phys.
\textbf{22} (1977) 1646, extended and translated from Ukrainian into English by authors }  }\\[0pt]
\vspace{20pt} {\large A.Yu. Korchin\footnote{E-mail:
korchin@kipt.kharkov.ua} and A.V. Shebeko\footnote{E-mail:
shebeko@kipt.kharkov.ua} \\[0pt] }
\vspace{8pt} {National Science Center ``Kharkov Institute of
Physics and
Technology'',\\[0pt] 61108 Kharkov, Ukraine}
\end{center}

\begin{abstract}
Analytical expressions for the elastic and dynamic form factors
(FFs) are derived in the shell model (SM) with a potential well of
finite depth. The consideration takes into account the motion of
the target-nucleus center of mass (CM). Explanation is suggested
for a simultaneous shrinking of the density and momentum
distributions of nucleons in nuclei. The convenient working
formulae are given to handle the expectation values of relevant
multiplicative operators in case of the $1s-1p$ shell nuclei.
\end{abstract}

\bigskip

{\bf 1}. It is known that nuclear SM wave functions (WFs)  do not
possess the property of translational invariance (TI). Several
methods in earlier and recent studies of nuclear structure (see,
e.g., $[1-4]$) have been proposed to transform any WF into one
which is translationally invariant. Of great interest among these methods is the projection
procedure considered in $[4]$. Along with other attractive features shown in $[4]$, this
procedure
enables a comparatively simple evaluation of the corresponding CM
correction to the purely shell quantities (see Ref.\,$[5]$).

The approach developed in $[5]$ is extended here to calculate the
cross sections of the elastic and quasifree electron scattering on
atomic nuclei with single-particle (s.p.) configurations more
complex than the $1s^4$ one. In particular, we pay special
attention to the physical interpretation of a simultaneous
shrinking of the density and momentum distributions of nucleons in
a nucleus due to the employed separation of its CM motion.

{\bf 2}. By definition, the elastic FF in question is
$$
F(\vec{q})=\langle \Phi_{intr}\mid \exp[\imath
\vec{q}(\hat{\vec{r}}_{1}-\hat{\vec{R}})] \mid \Phi_{intr}
\rangle, \eqno{(1)}
$$
while the dynamic FF can be written as in $[5]$,
$$
S(\vec{q},\omega)=\frac{1}{2\pi}\int\limits_{-\infty}^{\infty}
\exp(-\imath a \tau) S(\vec{q},\tau) d\tau, \eqno{(2)}
$$
$$
S(\vec{q},\tau)=\langle \Phi_{intr}\mid
\exp[\imath(\hat{\vec{p}}_{1}-A^{-1}\hat{\vec{P}})\vec{q}m^{-1}\tau]\mid
\Phi_{intr}\rangle,
$$
$$
a=\omega+q^{2}/{2m},
$$
where $\omega(\vec{q})$ is the energy (momentum) transfer, $m$ the nucleon mass,
$$
\hat{\vec{R}}=A^{-1}\sum\limits_{\alpha=1}^{A}
\hat{\vec{r}}_{\alpha} \qquad (\hat{\vec{P}}=
\sum\limits_{\alpha=1}^{A}\hat{\vec{p}}_{\alpha})
$$
the CM coordinate (total momentum) operator of the nucleus
composed of $A$ nucleons so that
$\hat{\vec{r}}_{\alpha}(\hat{\vec{p}}_{\alpha})$ the coordinate
(momentum) operator for nucleon number $\alpha$, and $\Phi_{intr}$
the intrinsic WF of the nuclear ground state (g.s.).

Following $[4]$, we take as $\mid \Phi_{intr}\rangle$ the vector
$$
\mid
\Phi_{intr}\rangle=(\vec{R}=0\mid\Phi\rangle[\langle\Phi\mid\vec{R}=0)
(\vec{R}=0\mid\Phi\rangle]^{-1/2},
\eqno{(3)}
$$
for a given trial (approximate) WF $\Phi$ that may be
nontranslationally invariant (nTI). Here, a round bracket,\,$\mid\, )$, is
used to represent a vector in the space of the CM coordinate only.

In the harmonic oscillator (HO) model, where the Slater
determinant $\mid \Phi \rangle$ is "pure" in the space of the CM
coordinate (the Bethe-Rose-Elliot-Skyrme theorem $[6,7]$), one has
$$
F(q)=\exp(\frac{q^{2}r_0^2}{4A})F_0(q) \eqno{(4)}
$$
with
$$
F_0(q)=\langle\Phi\mid \exp(\imath
\vec{q}\hat{\vec{r}}_{1})\mid\Phi \rangle \eqno{(4')}
$$
and
$$
S(\vec{q},\tau)=\exp(\frac{b^2\tau^2}{4A})S_0(\vec{q},\tau)
\eqno{(5)}
$$
with
$$
S_0(\vec{q},\tau)=\langle\Phi\mid
\exp(\imath\hat{\vec{p}}_1\vec{q}\tau/m)\mid \Phi \rangle
\eqno{(5')}
$$
Here $r_0$ is the oscillator parameter, $b=p_0q/m$, \,\, $p_0=r_0^{-1}$.

Result $(4)$ is widely used in applications starting from the work
$[8]$. Note also that Eqs.$(4)$ and $(5)$ are valid within the HO
model, being independent of any specific way to separate the CM
motion (e.g., the Ernst-Shakin-Thaler prescription (3) that is
equivalent to the so-called "fixed CM approximation" in case of
finite nuclei\,(bound systems) ).

Now, before finding some analogs of relations $(4)-(5)$  with an arbitrary WF\,
$\Phi$
(in particular, the Slater determinant constructed of the s.p. orbitals in a potential well
of finite depth, say, the Woods-Saxon or Hartree-Fock ones), we
would like to trace the CM corrections of the density and momentum
distributions $\rho(r)$ and $\eta(p)$ within the HO model. In this
connection, let us recall the general definitions for these quantities of primary  concern:
$$
\rho(\vec{r}) \equiv \langle\Phi_{intr}\mid \delta(\hat{\vec{r}}_{1}-\hat{\vec{R}}-\vec{r})\mid
\Phi_{intr}\rangle=
(2\pi)^{-3}\int\exp(-\imath\vec{q}\vec{r})F(\vec{q})d\vec{q}
\eqno{(6a)}
$$
and
$$
\eta(\vec{p})\equiv\langle \Phi_{intr}\mid
\delta(\hat{\vec{p}}_{1} - \hat{\vec{P}}/A-\vec{p})\mid
\Phi_{intr}\rangle=
$$
$$
=(2\pi)^{-3}(\tau/m)^{3}\int\exp(-\imath\vec{p}\vec{q}\tau/m)S(\vec{q},\tau)d\vec{q}
\eqno{(6b)}
$$
For the $1s$-$1p$ shell nuclei we find in the HO model,
$$
F_0(\vec{q})=(1-\frac{A-4}{6A}q^2r_0^2)\exp(-\frac14q^2r_0^2) ,
$$
$$
S_0(\vec{q},\tau)= (1-\frac{A-4}{6A}b^2\tau^2)\exp(-\frac14b^2\tau^2)
$$

Substituting these expressions, respectively, into Eq.$(4)$ and
Eq. $(5)$ we get with the help of formulae $(6)$:
$$
\rho(r)=\pi^{-3/2}\bar{r}_0^{-3}(\frac{3}{A-1}+\frac23 \frac{A-4}{A-1} \frac{r^2}{\bar{r}_0^2})
\exp(-\frac{r^2}{\bar{r}_{0}^2}),
\eqno{(7a)}
$$
$$
\eta(p)=\pi^{-3/2}\bar{p}_0^{-3}(\frac{3}{A-1}+\frac23 \frac{A-4}{A-1} \frac{p^2}{\bar{p}_0^2})
\exp(-\frac{p^2}{\bar{p}_{0}^2}),
\eqno{(7b)}
$$
$$
\bar{r}_0=\sqrt{\frac{A-1}{A}}r_0,~~
\bar{p}_0=\sqrt{\frac{A-1}{A}}p_0
$$

The intrinsic distributions without any CM correction are
$$
\rho_0(r)\equiv\langle \Phi\mid
\delta(\hat{\vec{r}}_1-\vec{r})\mid\Phi\rangle=\pi^{-3/2}r_0^{-3}(\frac4A+\frac{2(A-4)r^2}{3Ar_0^2})
\exp(-\frac{r^2}{r_0^2}),
\eqno{(8a)}
$$
$$
\eta_0(p)\equiv\langle \Phi\mid
\delta(\hat{\vec{p}}_1-\vec{p})\mid\Phi\rangle=\pi^{-3/2}p_0^{-3}(\frac4A+\frac{2(A-4)p^2}
{3Ap_0^2})\exp(-\frac{p^2}{p_0^2}),
\eqno{(8b)}
$$
By comparing Eqs.$(7)$ and $(8)$ one can see that the density and momentum distributions are
subject to the equal changes. It does not seem to be surprising if we invoke the well known
symmetry between the coordinate and momentum representations of the HO
model Hamiltonian. One should point out that, apart from the "most
symmetrical" nucleus with $A=4$, these CM corrections are not
reduced to a simple renormalization of the oscillator parameter
(e.g., the change $r_0\rightarrow\bar{r}_0$).

The dependences $\rho(r)$ and $\rho(r)_0$ calculated by formulae
$(7)$ and $(8)$  with $A=16$ are depicted in Fig.$1$  In this
context, we show the ratio
$$
\frac{\rho_0(r)}{\rho_0(0)}=[1+\frac16(A-4)\frac{r^2}{r_0^2}]\exp(-\frac{r^2}{r_0^2})
$$
versus the ratio
$$
\frac{\rho(r)}{\rho_0(r)}=\frac34[\frac{A}{A-1}]^{5/2}[1+\frac29(A-4)\frac{r^2}{\bar{r}_0^2}]
\exp(-\frac{r^2}{\bar{r}_0^2})
$$

\begin{figure}[t]
\begin{center}
\epsfig{file=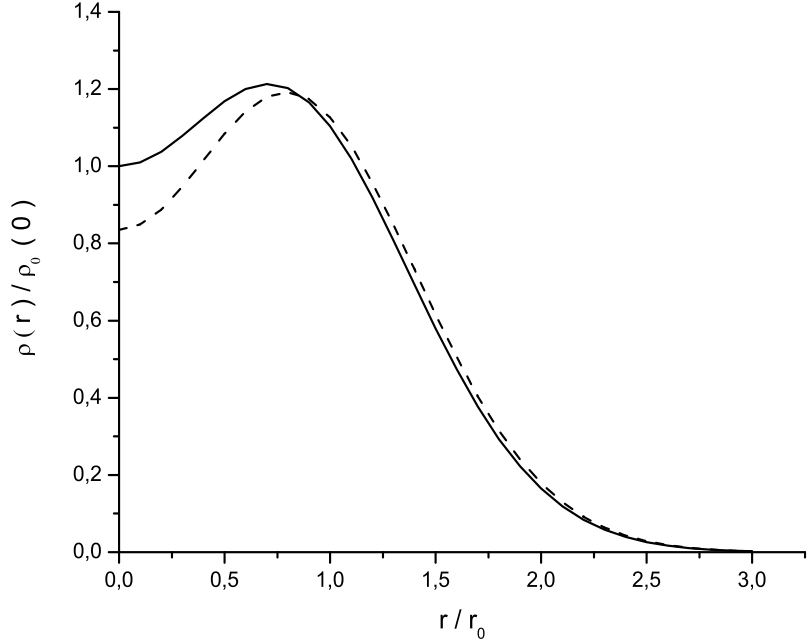,width=12cm,height=10cm} \caption{The
variation of $\rho_0(r)$/$\rho_0(0)$ (solid) and of
$\rho(r)$/$\rho_0(0)$ (dashed) with $r/r_0$}
\end{center}
\end{figure}

We see that an additional correlation between the nucleons, incorporated into (inherent in)
the intrinsic density $\rho(r)$, gives rise to their redistribution between the nuclear shells
(from the $1s$-shell to the $1p$-one, while the center of the
$1p$-distribution is shifted toward larger $r$-values). This
rearrangement of the nuclear interior is accompanied by a decrease
of the nuclear density in its peripheral region. It implies the
corresponding increase of the probability to find the nucleons in the central part of the nucleus.
Remind that both the density distributions (DDs) $\rho(\vec{r})$ and $\rho_0(\vec{r})$ are
normalized to unity. We interpret these features of the intrinsic
DD $\rho(r)$ as its shrinking  in comparison with the DD $\rho_0(r)$ , which embodies the spurious
CM motion modes. The same interpretation can be applied to the momentum distribution (MD)
$\eta(p)$ vs $\eta_0(p)$.

Such a simultaneous shrinking of the DD and MD becomes more
tractable if one evaluate the respective r.m.s. radii and momenta.
By definition, one has
$$
\langle r^2\rangle\equiv\int r^2 \rho(\vec{r}) d\vec{r},~~\langle
p^2\rangle\equiv\int p^2 \eta(\vec{p}) d\vec{p} \eqno{(9)}
$$
It is readily seen that
$$
\langle r^2\rangle_{0} \equiv\int r^2 \rho_0(\vec{r})
d\vec{r}=\langle r^2\rangle+(000\mid {\hat{\vec{R}} }^2\mid 000), \eqno{(10a)}
$$
$$
\langle p^2\rangle_{0} \equiv\int p^2 \eta_0(\vec{p})
d\vec{p}=\langle p^2\rangle+A^{-2}(000\mid {\hat{\vec{P}} }^2 \mid 000),
\eqno{(10b)}
$$
where the vector $\mid 000)$ is used to describe  the "zero"
(lowest-energy) CM oscillations with respect to the origin of
coordinates. A complementary  smearing of $\rho_0(r)$
and $\eta_0(p)$ compared respectively  with $\rho(r)$ and $\eta(p)$
is due to the nonphysical motion mode.

Further, let us consider the commutation relations for the relative coordinates
$\hat{\vec{r}}~'_\alpha=\hat{\vec{r}}_\alpha-\hat{\vec{R}}$ and
the canonically conjugate momenta
$\hat{\vec{p}}~'_\alpha=\hat{\vec{p}}_\alpha-\hat{\vec{P}}/A$
$(\alpha=1,\dots,A)$:
$$
[(\hat{\vec{r}}~'_\alpha)_k,(\hat{\vec{p}}~'_\alpha)_l]=\imath
\delta_{kl}(1-A^{-1}) ~(k,l=1,2,3) \eqno{(11)}
$$
along with the original ones:
$$
[(\hat{\vec{r}}_\alpha)_k,(\hat{\vec{p}}_\alpha)_l]=\imath
\delta_{kl}~(k,l=1,2,3) \eqno{(11')}
$$
The corresponding uncertainties meet the equations (see, for
instance, [9], p.\,54, and also Suppl.\,$A$ to this translation ) :
$$
\langle \Phi \mid (\widehat{\Delta \vec{r} })^2 \mid \Phi \rangle
\langle \Phi \mid (\widehat{\Delta \vec{p} })^2 \mid \Phi \rangle = \langle r^2 \rangle _0
\langle p^2\rangle _0 \geq \frac94, \eqno (12)
$$
$$
\langle \Phi \mid (\widehat{\Delta \vec{r}}~')^2 \mid \Phi \rangle
\langle \Phi \mid (\widehat{\Delta \vec{r}}~')^2 \mid \Phi \rangle =
{\langle r^2 \rangle}_{\Phi} { \langle p^2 \rangle}_{\Phi}
\geq \frac94 \left(\frac{A-1}{A} \right)^2 \eqno (12')
$$
for any state $\mid \Phi \rangle$ normalized to unity, where we
have introduced the expectation values $ {\langle r^2 \rangle}_{\Phi} =
\langle \Phi \mid (\hat{\vec{r}}_1')^2 \mid \Phi \rangle$ and
${\langle p^2 \rangle}_{\Phi} = \langle \Phi \mid (\hat{\vec{p}}_1')^2
\mid \Phi \rangle$. The latter are converted into the values
$\langle r^2 \rangle$  and $\langle p^2 \rangle$ if $\mid \Phi \rangle=\mid\Phi_{intr}\rangle$.

Alternatively, according Eqs.$(10)$ we find in the HO model that
$$
\langle r^2 \rangle \langle p^2 \rangle = \langle r^2 \rangle _0
\langle p^2\rangle _0 \left[1-\frac{(000\mid {\hat{\vec{R}} }^2\mid 000)}
{\langle r^2\rangle_0}\right]\left[1-\frac{(000\mid {\hat{\vec{P}} }^2 \mid 000)}
{A^2\langle p^2\rangle_0}\right] \eqno{(13)}
$$
Then, taking into account that
$$
(000\mid {\hat{\vec{R}} }^2\mid 000)=\frac{3}{2A}r_0^2,~~(000\mid {\hat{\vec{P}} }^2 \mid 000)=
\frac{3A}{2}p_0^2
$$
and that in the HO model for a nucleus with fully occupied (closed) shells
$$
\langle r^2 \rangle_0=\varkappa r_0^2, ~ \langle p^2 \rangle_0=\varkappa p_0^2,
$$
$$
\varkappa = 2 \sum\limits_{N=0}^{N_{max}}(N+1)(N+2)(N+3/2)/A,~~
A=2\sum\limits_{N=0}^{N_{max}}(N+1)(N+2), \eqno{(14)}
$$
i.e.,
$$
\varkappa=\frac32+\sum\limits_{N}N(N+1)(N+2)/\sum\limits_{N}(N+1)(N+2),
$$
where N is the principal quantum number, we get
$$
{\langle r^2 \rangle}_0 {\langle p^2 \rangle}_0 = \varkappa^2 \geq \frac94
\eqno{(15)}
$$
$$
\langle r^2 \rangle \langle p^2 \rangle= \varkappa^2
\left(\frac{A-1}{A}\right)^2\left[1+\frac{1}{A-1}\frac{\kappa-3/2}{\kappa}\right]^2\geq
\frac94 \left(\frac{A-1}{A}\right)^2 \eqno{(15')}
$$
Thus we arrive again to relations $(12)$ and $(12')$. From this
derivation it follows that Eqs. $(15)$ and $(15')$ without the
signs of inequality are permitted only for the $1s^4$
configuration in the HO model. Such a minimization of the
uncertainty relations is retained after making the CM correction.

{\bf 3}. Further, using the algebraic procedure applied in [$5$] for the
calculation of the expectation values $(1)$ and $(2)$ with the
Slater determinant $| \Phi \rangle= | (1s)^4 \rangle$ of the
simple HO orbitals, we find
$$
F(\vec{q})=\exp(-\frac{q^2\bar{r_0}^2}{4})U(\vec{q})/U(0),
\eqno{(16)}
$$
$$
U(\vec{q})=\int\exp(-\frac{\lambda^2
r_0^2}{4A})f(\vec{\lambda},\vec{q})d\vec{\lambda},
$$
$$
f(\vec{\lambda},\vec{q})=\langle\Phi\mid
\hat{O}_1(\vec{b})\prod\limits_{\alpha=2}^{A}
\hat{O}_{\alpha}(\vec{c})\mid \Phi \rangle
$$
with the operators
$$
\hat{O}_{\alpha}(\vec{x})= \exp(-\vec{x}^{\ast}~\hat{\vec{a}}_{\alpha}^{\dagger})
\exp(-\vec{x}~\hat{\vec{a}}_\alpha),~(\alpha=1,\dots,A)
$$
where
$\vec{b}=\vec{c}+\imath\frac{r_0}{\sqrt{2}}\vec{q},~~\vec{c}=
\imath\frac{r_0}{\sqrt{2}A}(\vec{\lambda}-\vec{q})$, $\hat{\vec{a}}^{\dagger}(\hat{\vec{a}})$ is
the vector whose components are the creation (annihilation) operators for
oscillator quanta in the three different space directions.

Analogously, one can show that
$$
S(\vec{q},\tau)=\exp(-\frac{q^2\bar{p_0}^2}{4m}\tau^2)U(-\imath p_0^2\frac{\vec{q}}{m}\tau)/U(0)
\eqno{(17)}
$$

Now, let us assume that the many-body state $\mid \Phi \rangle$ is a
Slater determinant of s.p. states $\mid \phi_\gamma\rangle~
(\gamma=1,\dots,A)$ which describe completely occupied bound
states of the nucleons in a spherically symmetric potential well
(e.g., Woods-Saxon potential or Hartree-Fock field). Then (see Suppl.\,B)
$$
f(\vec{\lambda},\vec{q})=A^{-1}\sum\limits_{\rho=1}^A D_\rho,
\eqno{(18)}
$$
where $D_\rho$ is the determinant that is deduced from the determinant
$$
D=\parallel \langle\phi_\gamma\mid
\hat{O}(\vec{c})\mid\phi_{\gamma '}\rangle \parallel
$$
replacing the vector $\vec{c}$ by $\vec{b}$ in its column  with
the label $\rho$.

As an illustration, let us consider the $(1s)^4(1p)^{12}$
configuration in the $ls$-coupling scheme. From Eq.$(18)$ it
follows that (see Suppl.\,$C$)
$$
f(\vec{\lambda},\vec{q})=\frac14d^3(d_1+d_2+d_3+d_4), \eqno{(19)}
$$
where
$d=\parallel \langle\phi_{nlm}\mid\hat{O}(\vec{c})\mid\phi_{n'l'm'}\rangle \parallel$
is the $4\times4$ determinant, $\mid\phi_{nlm}\rangle$ the spatial
part of the vector $\mid \phi_\gamma\rangle$ in the shell with
the radial quantum number $n$, the orbital angular momentum $l$
and its projection $m$, and $d_i(i=1,\dots,4)$ are deduced from
$d$ just as $D_\rho$ from $D$.

Subsequent simplifications can be achieved owing to the transformation properties of the matrix
elements
$$
M_{nlm}^{n'l'm'}(\vec{x})\equiv\langle\phi_{nlm}\mid\hat{O}(\vec{x})\mid\phi_{n'l'm'}\rangle
$$
with respect to the rotation group. In fact, we have
$$
M_{1s}^{1s}(\vec{x})=M_0(x^2), \qquad M_{1pm}^{1s}(\vec{x})=M(x^2)x_m^\ast,
$$
$$
M_{1pm}^{1pm'}(\vec{x})=M_1(x^2)\delta_{mm'}+M_2(x^2)x_m^\ast
x_{m'}~(m,m'=-1,0,1) \eqno{(20)}
$$
with $\vec{x}^\ast=-\vec{x}$. Here $x_m$ are the spherical
components of the vector $\vec{x}$.

In their turn, the determinants $d_i$ can be expressed in terms of
the scalar functions $M_0,M,M_1$ and $M_2$ if one takes into
account that the quantities $d_1$ and $d_2+d_3+d_4$, each
separately, are invariant under rotations, i.e., they depend on
$b^2,c^2$ and $\vec{b}\vec{c}$. This property enables us to write
down,
$$
d_1=M_1^2(c^2)\{M_0(b^2)N_0(c^2)-M(b^2)M(c^2)\vec{b}\vec{c}\},
\eqno{(21)}
$$
$$
d_2+d_3+d_4=M_1(c^2)\{M_1(b^2)M_0(c^2)-M(b^2)M(c^2)\vec{b}\vec{c}\}+
$$
$$
2M_1(b^2)M_1(c^2)\{M_0(c^2)N_0(c^2)-M^2(c^2)c^2\}+
$$
$$
+M_0(c^2)M_1^2(c^2)M_2(b^2)b^2-M_1(c^2)M_2(b^2)\{M_2(c^2)+M(c^2)\}\{b^2c^2-
(\vec{b}\vec{c})^2\}
\eqno{(22)}
$$
with $N_0(x^2)=M_{1p0}^{1p0}(x\vec{e}_0)$, where $\vec{e}_0$ is the unit vector along the Z-axis.

The scalars $M(x^2)$ and $M_1(x^2)$ satisfy the relations
$$
xM(x^2)=M_{1s}^{1p0}(x\vec{e}_0), \eqno{(23)}
$$
$$
M_1(x^2)=\frac12\{M_{1p,-1}^{1p,-1}(x\vec{e}_0)+M_{1p,+1}^{1p,+1}(x\vec{e}_0)\},
\eqno{(24)}
$$
from which it follows that
$$
M_1(x^2)=\exp(r_0^2y^2/4)(A_0(y)+A_2(y)), \eqno{(25)}
$$
$$
x^2M_2(x^2)=-3\exp(r_0^2y^2/4)A_2(y) \eqno{(26)}
$$
with
$$
A_{\lambda}(y)=\int\limits_0^{\infty}j_{\lambda}(yr)R_{11}(r)r^2dr,~(\lambda=0,2)
$$
where $R_{11}(r)$ is the radial part of $\phi_{11m}(\vec{r})$ and
$j_{\lambda}(z)$ is the spherical Bessel function of $z$. In Eqs.\,
$(25)-(26)$ $x=\imath r_0y/{\sqrt{2}}$.

Thus the initial cumbersome task of handling the expectation value
$f(\vec{\lambda},\vec{q})$ is reduced to calculation of the simple
overlap integrals.

All these formulae can be useful when studying CM corrections of
the cross sections of the elastic and quasifree electron
scattering from nuclei more complicated than the $^4He$ nucleus.
An attractive feature of similar studies is to proceed with one
and the same corrected many-body WF of the nuclear g.s. in
evaluating different structure functions like the DD and MD. Of
certain interest might be deviations of the $q$-dependence of the
ratio $F(q)/F_0(q)$ beyond the HO model from that which is given
by the canonical Tassie-Barker factor $\exp(q^2r_0^2/4A)$.

\section*{Acknowledgements}

\hspace{0.5cm} We would like to thank Pavel Grigorov for technical
assistance.


\begin{appendix}
\section*{Appendix A. Comments on simultaneous shrinking of density and momentum
distributions} \label{app:A} \setcounter{equation}{0}
\def\theequation{A.\arabic{equation}}

\bigskip

Remind that a commutation relation between two noncommuting
operators imposes a definite constraint upon their dispersions,
i.e., the mean square deviations of these quantities from the
corresponding expectation values. In fact, let us consider the two
Hermitean operators $\hat A$ and $\hat B$ that meet
$$
\left[ \hat A, \hat B \right] = \imath \hat C\ , \eqno({\rm A.1})
$$
where $\hat C$ is also an Hermitean operator.
In particular, if $\hat A = \hat x$ and $\hat B = {\hat p}_x$,
then the operator $ \hat C = \hbar $ \footnote{Recall that in our
system of units $\hbar = 1$}.

By definition, the expectation values with respect to an arbitrary
state $\Phi $ are
$$
\langle \hat A \rangle = \langle \Phi | \hat A | \Phi \rangle ,
\quad \langle \hat B \rangle = \langle \Phi | \hat B | \Phi
\rangle\ . \eqno({\rm A.2})
$$

Let us introduce the deviations
$$
\widehat{\Delta A} = \hat A - \langle \hat A \rangle , \quad
\widehat{\Delta B} = \hat B - \langle \hat B \rangle \ .
\eqno({\rm A.3})
$$
Obviously they satisfy the relation
$$
\left[ \widehat{\Delta A}, \widehat{\Delta B} \right] = \imath
\hat C\ . \eqno({\rm A.4})
$$

From Eq.(\rm A.4) it follows the uncertainty relation
$$
\langle (\widehat{\Delta A})^2 \rangle \langle (\widehat{\Delta
B})^2 \rangle \ge \frac{1}{4} {\langle \hat C \rangle}^2\ .
\eqno({\rm A.5})
$$
Note also that for operator $\hat X$,
$$
\langle (\widehat{\Delta X})^2 \rangle = \langle (\hat X - \langle
\hat X \rangle )^2 \rangle = \langle {\hat X}^2 \rangle - {\langle
\hat X \rangle}^2\ , \eqno({\rm A.6})
$$
if $\langle \Phi \mid \Phi \rangle = 1 $.

In the case of interest for the relative coordinates
${\hat{\ve{r}}}'_{\alpha}$ and the canonically conjugate momenta
${\hat{\ve{p}}}'_{\alpha}$ the corresponding relation looks as
$$
\langle {\Phi}_{int} \mid (\widehat{\Delta \ve{r'}})^2 \mid
{\Phi}_{int} \rangle \langle {\Phi}_{int} \mid (\widehat{\Delta
\ve{p'}})^2 \mid {\Phi}_{int} \rangle \ge \frac{9}{4} { \left(
\frac{A - 1}{A} \right) }^2\ , \eqno({\rm A.7a})
$$
or
$$
\langle r^2 \rangle \langle p^2 \rangle = 9 \langle x^2 \rangle
\langle {p_x}^2 \rangle \ge \frac{9}{4} {\left( \frac{A - 1}{A}
\right)}^2\  \eqno({\rm A.7b})
$$
for any state ${\Phi}_{int}$ normalized to \underline{unity}.
Here, in accordance with Eq.({\rm A.6}),
$$
\langle {\Phi}_{int} \mid (\widehat{\Delta \ve{r'}})^2 \mid
{\Phi}_{int} \rangle = \langle {\Phi}_{int} \mid (
{\hat{\ve{r}}_1}')^2 \mid {\Phi}_{int} \rangle \equiv \langle {
\ve{r}}^2 \rangle = \int r^2 \rho (r) {\rm d} \ve{r} \ \eqno({\rm
A.8})
$$
and
$$
\langle {\Phi}_{int} \mid (\widehat{\Delta \ve{p'}})^2 \mid
{\Phi}_{int} \rangle = \langle {\Phi}_{int} \mid ( {\hat{
\ve{p}}_1}')^2 \mid {\Phi}_{int} \rangle \equiv \langle {
\ve{p}}^2 \rangle = \int p^2 \eta (p) {\rm d} \ve{p} \ ,
\eqno({\rm A.9})
$$
since $\langle \hat{\ve{r'}} \rangle = \langle \hat{\ve{p'}}
\rangle = 0$.

Thus, the general result ({\rm A.5}) leads to the condition ({\rm
A.7b}) for the pair $ {\hat{\ve{r_1}}}' $   and $
{\hat{\ve{p_1}}}' $ that obeys the commutation rules,
$$
\left[ ({\hat{\ve{r}}_1}')_j, ({\hat{\ve{p}}_1}')_k \right] =
\imath {\delta}_{jk} \frac{A-1}{A } \quad (j, k = 1,2,3) . \
\eqno({\rm A.10})
$$
Eqs.({\rm A.10}) are different from
$$
\left[ (\hat{\ve{r}}_1)_j, (\hat{\ve{p}}_1)_k \right] = \imath
{\delta}_{jk}  \quad (j, k = 1,2,3) . \  \eqno({\rm A.11})
$$
By the way, this fact means that the transformation
${\hat{\ve{r}}}_{\alpha} \to {\hat{\ve{r}}}'_{\alpha}$ and
${\hat{\ve{p}}}_{\alpha} \to {\hat{\ve{p}}}'_{\alpha}  \, (\alpha
= 1, \dots , A)$ is nonunitary, i.e., it cannot be performed via a
unitary operator $\hat U  \ ( {\hat U}^{\dag} \hat U = \hat U
\hat{U}^{\dag} = 1) $, viz., putting $ \hat{U}^{\dag}
{\hat{\ve{r}}}_{\alpha} \hat U = {\hat{\ve{r}}}'_{\alpha} $ and $
{\hat U}^{\dag} {\hat{\ve{p}}}_{\alpha} \hat U =
{\hat{\ve{p}}}'_{\alpha}$. Under the latter the commutation
relations would not change.

It follows from  ({\rm A.5}) and ({\rm A.11}) that
$$
{\langle r^2 \rangle}_0 { \langle p^2 \rangle}_0 \ge \frac{9}{4}\
\eqno({\rm A.12})
$$
for the expectation values
$$
{\langle r^2 \rangle}_0 \equiv \langle \Phi \mid
{\hat{\ve{r}}_1}^2 \mid \Phi \rangle = \int r^2 {\rho}_0 (r) {\rm
d} \ve{r} \ \eqno({\rm A.13})
$$
and
$$
{\langle p^2 \rangle}_0 \equiv \langle \Phi \mid
{\hat{\ve{p}}_1}^2 \mid \Phi \rangle = \int p^2 {\eta}_0 (p) {\rm
d} \ve{p} \  \eqno({\rm A.14})
$$
in the shell model ground state $\Phi $ (in general, any state
$\Phi $) such that $\langle \Phi \mid \Phi \rangle = 1$.

So, we have relation ({\rm A.7b}) for the mean square radius and
mean square momentum in an exact translationally invariant state
versus relation ({\rm A.12}) obtained for similar quantities
within an approximate treatment of the system (say, with the
wavefunction which has a violated symmetry under space
translations). In comparison with Eq.({\rm A.12}), the r.h.s. of
Eq.({\rm A.7b}) is modified by the factor ${\left( \frac{A - 1}{A}
\right)}^2 \le 1$. This modification is closely connected with
nonunitarity of the transformation from the usual coordinates and
canonically conjugate momenta to their relative counterparts.

Thus the simultaneous shrinking of the density and momentum
distributions, shown in the paper within the HOM, is consistent
with the model independent uncertainty relations ({\rm A.7}).

\end{appendix}

\begin{appendix}
\section*{Appendix B. Comments on derivation of Eq.(18)}
\label{app:B} \setcounter{equation}{0}
\def\theequation{B.\arabic{equation}}

One has to deal with the expectation values of type
$$
A_{\Phi}=\langle \Phi \mid \hat
{O}_{1}(\vec{b})\hat{O}_{2}(\vec{c})\dots \hat{O}_{A}(\vec{c})\mid
\Phi \rangle, \eqno({\rm B.1})
$$
$$
\hat{O}_{\gamma}(\vec{x})=\exp(-\vec{x}^\ast\hat{\vec{a}}^\dag_{\gamma})
\exp(\vec{x}\hat{\vec{a}}_{\gamma})\equiv\hat{E}^\dag_{\gamma}(-\vec{x})\hat{E}_{\gamma}(\vec{x}),
(\gamma=1,\dots,A) \eqno({\rm B.2})
$$
where, for instance, $\vec{b}=\vec{c}+\vec{s}$. One can write
$$
A_{\Phi}=\langle \Phi \mid
\hat{E}_{1}^\dag(-\vec{c})\hat{E}_{2}^\dag(-\vec{c})\dots
\hat{E}_{A}^\dag(-\vec{c})\hat{E}_{1}^\dag(-\vec{s})\hat{E}_{1}(\vec{s})\hat{E}_{1}(\vec{c})
\hat{E}_{2}(\vec{c})\dots\hat{E}_{A}(\vec{c})\mid \Phi \rangle .
\eqno({\rm B.3})
$$
We have used the properties
$E_{1}(\vec{c}+\vec{s})=E_{1}(\vec{c})E_{1}(\vec{s})$ and
$[E_{\alpha}(\vec{x}), E_{\beta}(\vec{y})]=0$ for any vectors
$\vec{x}$ and $\vec{y}$.

If $\mid \Phi \rangle$ is a Slater determinant, i.e.,
$$
\mid \Phi \rangle= \mid Det \rangle= \sqrt{A!}\hat{\Omega}\mid
\phi_{1}(1)\rangle\dots \mid\phi_{A}(A)\rangle \eqno({\rm B.4})
$$
with the antisymmetrization operator
$$
\hat{\Omega}=(A!)^{-1}\sum\limits_{P} \varepsilon_{p}\hat{P}
\eqno({\rm B.5})
$$
which has the property
$$
\hat{\Omega}^{2}=\hat{\Omega}, \eqno({\rm B.6})
$$
then
$$
A_{\Phi}=\langle Det(-\vec{c})\mid
\hat{E}_{1}(-\vec{s})\hat{E}_{1}(\vec{s})\mid Det(\vec{c}) \rangle
\eqno({\rm B.7})
$$
with
$$
\mid Det(\vec{c})\rangle=\hat{E}_{1}(\vec{c})\dots
\hat{E}_{A}(\vec{c}) \mid Det \rangle=
\sqrt{A!}\hat{\Omega}\hat{E}_{1}(\vec{c})\mid \phi_{1}(1)
\rangle\dots\hat{E}_{A}(\vec{c})\mid \phi_{A}(A)\rangle \eqno({\rm
B.8})
$$
Furthermore, using the permutation symmetry of the determinants
involved, viz.,
$$
\hat{P} \mid Det(-\vec{c}) \rangle= \varepsilon_{p} \mid
Det(-\vec{c}) \rangle \eqno({\rm B.9a})
$$
$$
\hat{P} \mid Det(\vec{c}) \rangle= \varepsilon_{p} \mid
Det(\vec{c}) \rangle , \eqno({\rm B.9b})
$$
it is easily seen that
$$
A_{\Phi}=\langle Det(-\vec{c})\mid
\hat{E}_{1}(-\vec{s})\hat{E}_{1}(\vec{s})\mid Det(\vec{c}) \rangle
$$
$$
=\langle Det(-\vec{c})\mid
\hat{E}_{2}(-\vec{s})\hat{E}_{2}(\vec{s})\mid Det(\vec{c})
\rangle=\dots
$$
$$
=\frac1A \langle Det(-\vec{c})\mid \sum\limits_{\alpha=1}^A
\hat{E}_{\alpha}(-\vec{s})\hat{E}_{\alpha}(\vec{s})\mid
Det(\vec{c}) \rangle
$$
Now, taking into account Eq.$(B.6)$ and the relation
$$
[\hat{\Omega},\sum\limits_{\alpha=1}^{A}\hat{B}_{\alpha}]=0
$$
for any A operators $\hat{B}_{1},\hat{B}_{2},\dots, \hat{B}_{A}$,
we find
$$
A_{\Phi}=\frac1A \langle \psi'_{1}(1)\mid \langle \psi'_{2}(2)\mid
\dots \langle \psi'_{A}(A)\mid \sum\limits_{\alpha=1}^A
E_{\alpha}^{\dag}(-\vec{s})E_{\alpha}(\vec{s})
\sum\limits_{p}\varepsilon_{p}\hat{P} \mid \psi_{1}(1)\rangle \mid
\psi_{2}(2)\rangle \dots \mid \psi_{A}(A)\rangle \eqno({\rm B.10})
$$
where we have introduced the two sets $\{\psi \}$ and $\{\psi' \}$
of new orbitals
$$
\mid \psi(\alpha)\rangle = \hat{E}_{\alpha}(\vec{c}) \mid
\phi(\alpha)\rangle \eqno({\rm B.11a})
$$
and
$$
\mid \psi'(\alpha)\rangle = \hat{E}_{\alpha}(-\vec{c}) \mid
\phi(\alpha)\rangle ~~ (\alpha=1,\dots, A) . \eqno({\rm B.11b})
$$
Expression (B.10) explicitly reads
$$
A_{\Phi}=\frac1A ~[~\sum\limits_{P}\varepsilon_{p}\langle
\psi'_{1}(1)\mid\hat{E}^\dag(-\vec{s})\hat{E}(\vec{s})\mid \psi
_{p_1}(1)\rangle\langle\psi'_2\mid \psi_{p_2}\rangle \dots \langle
\psi'_A \mid \psi_{P_A}\rangle+
$$
$$
+\sum\limits_{P}\varepsilon_{p}\langle \psi'_1 \mid
\psi_{P_1}\rangle \langle
\psi'_{2}(2)\mid\hat{E}^\dag(-\vec{s})\hat{E}(\vec{s})\mid \psi
_{p_2}(2)\rangle\dots \langle \psi'_A \mid \psi_{P_A}\rangle+
$$
$$
+\sum\limits_{P}\varepsilon_{p}\langle \psi'_1 \mid
\psi_{P_1}\rangle \langle \psi'_1 \mid \psi_{P_1}\rangle \dots
\langle \psi'_{A}(A)\mid\hat{E}^\dag(-\vec{s})\hat{E}(\vec{s})\mid
\psi _{p_A}(A)\rangle ~].
$$
The latter is equivalent to Eq.(18).
\end{appendix}

\begin{appendix}
\section*{Appendix C. Evaluation of determinants in Eq.(18)}
\label{app:C} \setcounter{equation}{0}
\def\theequation{C.\arabic{equation}}

In order to simplify evaluation of the $(1s)^{4}(1p)^{12}$
configuration determinants involved in the r.h.s. of Eq.$(18)$,
let us consider a sparse {\bf nm} $\times$ {\bf nm} matrix
$$Z(nm\times nm)= \begin{bmatrix} Z_{11}(m\times m) &
Z_{12}(m\times m) & \hdots \
Z_{1n}(m\times m)\\
Z_{21}(m\times m) & Z_{22}(m\times m) & \hdots \
Z_{2n}(m\times m)\\
\hdots & \hdots & \hdots\\
Z_{n1}(m\times m) & Z_{n2}(m\times m) & \hdots \ Z_{nn}(m\times m)\\
\end{bmatrix}
\eqno{(\rm C.1)}$$ that consists of $ {\bf n}^2$ diagonal
\textbf{m}$\times$\textbf{m} block matrices
$$
Z_{ik}(m\times m)\equiv {\rm diag}[
Z_{ik}^{(1)},Z_{ik}^{(2)},\dots,Z_{ik}^{(m)}] \eqno{(\rm C.2a)}
$$
or
$$Z_{ik}(m \times m)=\begin{bmatrix} \bullet &
0 & \dots \
0\\
0 & \di & \dots \
0\\
\dots & \hdots & \hdots\\
0 & 0 & \hdots \ \di \\
\end{bmatrix}
\eqno{(\rm C.2b)}$$
\begin {center}
$(i,k=1,2,\dots,n)$.
\end{center}
In the representation (C.2b) the diagonal elements $Z_{ik}^{(l)}$
(l=1,2,\dots,m) which are, in general, arbitrary and different
from one another, are marked by the symbol $\di$. One can show
that $det\ Z = \parallel Z(nm\times nm) \parallel$ is equal to the
product
$$
det\ Z= \prod_{i=1}^m det\ Y_i  \eqno{(\rm C.3)}
$$
of the \textbf{m} determinants of the \textbf{n}$\times
$\textbf{n} matrices $Y_i~(i=1, \dots, m)$. Each of $Y_i$ is
composed of the $\textbf{n}^2$ diagonal elements of the matrices
$Z_{ik}(m \times m) $ (namely, elements $\di$), which one
encounters when passing in the clockwise (or counter-clockwise)
direction the corresponding spiral-like contour as displayed in
Fig.~2.

One should note that these \textbf{m} contours are not closed
(i.e., they are open), and the $i$-th contour begins at the
element $Z_{11}^{(i)}$ on the diagonal of the block $Z_{11}$.

The relationship (C.3) can be proved using the simple properties
of determinants. As an illustration, let us demonstrate relation
(C.3) for $m=2$ and $n=3$:
$$~~~~~~~~~\begin{Vmatrix}
a_1 & 0 & b_1 & 0 & e_1 & 0\\
0 & a_2 & 0 & b_2 & 0 & e_2\\
c_1 & 0 & d_1 & 0 & f_1 & 0\\
0 & c_2 & 0 & d_2 & 0 & f_2\\
g_1 & 0 & h_1 & 0 & j_1 & 0\\
0 & g_2 & 0 & h_2 & 0 & j_2\\
\end{Vmatrix}= -\begin{Vmatrix}
a_1 & b_1 & 0 & 0 & e_1 & 0\\
0 & 0 & a_2 & b_2 & 0 & e_2\\
c_1 & d_1 & 0 & 0 & f_1 & 0\\
0 & 0 & c_2 & d_2 & 0 & f_2\\
g_1 & h_1 & 0 & 0 & j_1 & 0\\
0 & 0 & g_2 & h_2 & 0 & j_2\\
\end{Vmatrix}=
$$
$$
~~~~~~=\begin{Vmatrix}
a_1 & b_1 & 0 & 0 & e_1 & 0\\
c_1 & d_1 & 0 & 0 & f_1 & 0\\
0 & 0 & a_2 & b_2 & 0 & e_2\\
0 & 0 & c_2 & d_2 & 0 & f_2\\
g_1 & h_1 & 0 & 0 & j_1 & 0\\
0 & 0 & g_2 & h_2 & 0 & j_2\\
\end{Vmatrix}=
-\begin{Vmatrix}
a_1 & b_1 & e_1 & 0 & 0 & 0\\
c_1 & d_1 & f_1 & 0 & 0 & 0\\
0 & 0 & 0 & b_2 & a_2 & e_2\\
0 & 0 & 0 & d_2 & c_2 & f_2\\
g_1 & h_1 & j_1 & 0 & 0 & 0\\
0 & 0 & 0 & h_2 & g_2 & j_2\\
\end{Vmatrix}=
$$
$$
=-\begin{Vmatrix}
a_1 & b_1 & e_1 & 0 & 0 & 0\\
c_1 & d_1 & f_1 & 0 & 0 & 0\\
g_1 & h_1 & j_1 & 0 & 0 & 0\\
0 & 0 & 0 & b_2 & a_2 & e_2\\
0 & 0 & 0 & d_2 & c_2 & f_2\\
0 & 0 & 0 & h_2 & g_2 & j_2\\
\end{Vmatrix}=
~~\begin{Vmatrix}
a_1 & b_1 & e_1 & 0 & 0 & 0\\
c_1 & d_1 & f_1 & 0 & 0 & 0\\
g_1 & h_1 & j_1 & 0 & 0 & 0\\
0 & 0 & 0 & a_2 & b_2 & e_2\\
0 & 0 & 0 & c_2 & d_2 & f_2\\
0 & 0 & 0 & g_2 & h_2 & j_2\\
\end{Vmatrix} .
$$
Therefore
$$\begin{Vmatrix}
a_1 & 0 & b_1 & 0 & e_1 & 0\\
0 & a_2 & 0 & b_2 & 0 & e_2\\
c_1 & 0 & d_1 & 0 & f_1 & 0\\
0 & c_2 & 0 & d_2 & 0 & f_2\\
g_1 & 0 & h_1 & 0 & j_1 & 0\\
0 & g_2 & 0 & h_2 & 0 & j_2\\
\end{Vmatrix}=
\begin{Vmatrix}
a_1 & b_1 & e_1 \\
c_1 & d_1 & f_1 \\
g_1 & h_1 & j_1 \\
\end{Vmatrix}\cdot
\begin{Vmatrix}
a_2 & b_2 & e_2 \\
c_2 & d_2 & f_2 \\
g_2 & h_2 & j_2 \\
\end{Vmatrix} .
$$
The aforementioned prescription yields the same result but much
quicker.

\begin{figure}[t]
\begin{center}
\epsfig{file=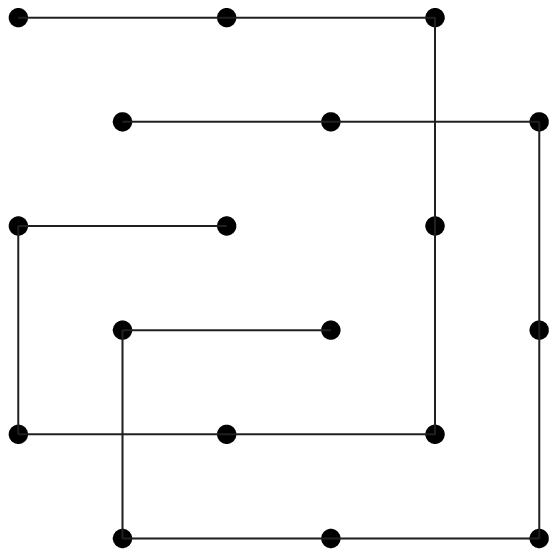,width=5cm,height=5cm} \caption{The
relevant contours in the case with $m=2$ and $n=3$.}
\end{center}
\end{figure}

In the case of interest one has to deal with the matrices of the
type
 $$M=\begin{bmatrix} M_{ss} & M_{s+} & M_{s0} & M_{s-}\\
M_{+s} & M_{++} & M_{+0} & M_{+-}\\
M_{0s} & M_{0+} & M_{00} & M_{0-}\\
M_{-s} & M_{-+} & M_{-0} & M_{--}\\
\end{bmatrix} ,
$$
where each of the sixteen $4 \times 4$ blocks is diagonal, e.g.,
$$
M_{ss}={\rm diag}[M_{ss}^1,M_{ss}^2,M_{ss}^3,M_{ss}^4]
$$
with the four equivalent dispositions:
$$
{\rm
diag}[M_{1s}^{1s}(\vec{b}),M_{1s}^{1s}(\vec{c}),M_{1s}^{1s}(\vec{c}),M_{1s}^{1s}(\vec{c})
] ,
$$
$$
{\rm
diag}[M_{1s}^{1s}(\vec{c}),M_{1s}^{1s}(\vec{b}),M_{1s}^{1s}(\vec{c}),M_{1s}^{1s}(\vec{c})]
,
$$
$$
{\rm
diag}[M_{1s}^{1s}(\vec{c}),M_{1s}^{1s}(\vec{c}),M_{1s}^{1s}(\vec{b}),M_{1s}^{1s}(\vec{c})]
,
$$
$$
{\rm
diag}[M_{1s}^{1s}(\vec{c}),M_{1s}^{1s}(\vec{c}),M_{1s}^{1s}(\vec{c}),M_{1s}^{1s}(\vec{b})]
.
$$
Taking into account this equivalence we get expression (19) which
is much simpler compared to Eq.(18).
\end{appendix}

\end{document}